# Buried and accessible surface area control intrinsic protein flexibility

Joseph A. Marsh

*European Bioinformatics Institute (EMBL-EBI)*

*Wellcome Trust Genome Campus*

*Hinxton, Cambridge CB10 1SD, United Kingdom*

*jmarsh@ebi.ac.uk*

Keywords: protein dynamics; protein structure; protein folding; monomer; solvent accessible surface area

# Abstract

Proteins experience a wide variety of conformational dynamics that can be crucial for facilitating their diverse functions. How is the intrinsic flexibility required for these motions encoded in their three-dimensional structures? Here, the overall flexibility of a protein is demonstrated to be tightly coupled to the total amount of surface area buried within its fold. A simple proxy for this, the relative solvent accessible surface area ($A_{rel}$), therefore shows excellent agreement with independent measures of global protein flexibility derived from various experimental and computational methods. Application of $A_{rel}$ on a large scale demonstrates its utility by revealing unique sequence and structural properties associated with intrinsic flexibility. In particular, flexibility as measured by $A_{rel}$ shows little correspondence with intrinsic disorder, but instead tends to be associated with multiple domains and increased $\alpha$-helical structure. Furthermore, the apparent flexibility of monomeric proteins is found to be useful for identifying quaternary structure errors in published crystal structures. There is also a strong tendency for the crystal structures of more flexible proteins to be solved to lower resolutions. Finally, local solvent accessibility is shown to be a primary determinant of local residue flexibility. Overall this work provides both fundamental mechanistic insight into the origin of protein flexibility and a simple, practical method for predicting flexibility from protein structures.

# Introduction

Proteins are intrinsically flexible, dynamic molecules. Although the structure-function insight obtained from the tens of thousands of X-ray crystal structures determined to date has demonstrated the tremendous utility of simple single-conformation models, it has long been clear, from the basic principles of statistical thermodynamics, that an ensemble representation would be required to fully describe a protein in solution. That is, rather than adopting unique structures, proteins can be considered as ensembles of multiple distinct conformers. A large body of experimental, theoretical and computational work now supports the importance of this energy landscape paradigm[1–3].

Although the ensemble view of protein structure is now firmly established, much progress is currently being made in characterizing the diverse ways that proteins populate the energy landscape. For many proteins, the conformational fluctuations are small and the classical representation of proteins as single unique structures is adequate for most practical purposes. Furthermore, the very fact that many proteins can be crystallized, and that those crystals are densely packed[4], provides strong justification for the approximation of many proteins as rigid solids. However, beyond this, it is clear that proteins undergo a wide range of motions that can be important for their functions. These can involve relatively minor backbone or side-chain dynamics or larger scale movements of secondary structural elements or domains[5]. In some cases, proteins can even be intrinsically disordered, *i.e.* partially or completely unfolded in solution[6–8], in which case the ensemble representation must cover a vast range of conformational space[9,10]. Given this structural and dynamic diversity, protein flexibility can be best considered on a continuum, with rigid, globular proteins at one extreme and intrinsically disordered proteins at the other[11–13].

We recently introduced a simple parameter, the relative solvent accessible surface area ($A_{rel}$), which describes the amount of surface area a protein exposes to solution (and, conversely, how little it buries intramolecularly) compared to what is expected given its molecular weight. $A_{rel}$ was shown to have great utility for predicting the magnitude of conformational changes that occur upon binding from the structures of protein complexes, allowing the demonstration that large conformational changes are extremely common[14]. This work also hinted of a close relationship between $A_{rel}$ and protein flexibility, implying a tight correspondence between the intrinsic flexibility of proteins in their unbound states and their binding-induced conformational changes.

Here, the theoretical basis for a relationship between buried and accessible surface area and intrinsic protein flexibility is first discussed, and then validated by demonstrating strong correlations between $A_{rel}$ and various measures of flexibility calculated from normal mode analysis, NMR ensemble models, molecular dynamics simulations and NMR chemical shifts. This enables analyses of the associations between intrinsic flexibility and various protein properties, including domain structure, amino acid composition, secondary structure, quaternary structure and crystal-structure resolution. Finally, the relationship between solvent accessibility and local flexibility is investigated.

## Results and Discussion

### *Origins of the relations between molecular weight, solvent accessible surface area and intrinsic flexibility*

As the crystal structures of an increasing number of proteins were determined in the 1970s and 1980s, an interesting phenomenon was noted: when the solvent accessible surface areas of proteins was plotted against their molecular weights, a very tight correspondence was observed[15–17]. This still holds today with the large number of crystal structures that have now been determined, as shown in Figure 1A. From this plot, the expected solvent accessible surface area for a folded, crystallizable, monomeric protein ($A_{monomer}$) of molecular weight $M$ can be fit:

$$A_{monomer} = 4.44 M^{0.770} \quad (1)$$

At first glance, this relationship might be attributed to a simple geometric phenomenon in which surface area should scale with an exponent of 2/3 with respect to volume[16,17]. In fact, given the difference between molecular surface and solvent accessible surface area (*i.e.* the presence of a solvent layer), an exponent of slightly less than 2/3 might be expected[15]. However, as is seen here and noted previously[15,18], an exponent considerably higher than 2/3 provides the best fit. One simple interpretation of this is that larger proteins tend to adopt more extended conformations, burying less surface area per residue than if they adopted a constant shape with increasing size. Similarly, it has been suggested that bigger proteins are less densely packed[19,20]. Finally, an energetic explanation has been proposed: since protein folding is driven by the burial of surface area in order to compensate for the massive conformational entropy of the unfolded state[21–23], Equation 1 is influenced by this tendency to bury surface area in proportion to unfolded-state entropy, which in turn is proportional to molecular weight[18].

Regardless of its fundamental origin, one can easily accept Equation 1 as an empirical relationship describing what is expected for a typical monomeric, crystallizable protein of a given molecular weight. We recently showed that the relative solvent accessible surface area, $A_{rel}$, defined as the observed solvent accessible surface area, $A_{observed}$, divided by the expected value, $A_{monomer}$, could be extremely useful for understanding the changes in conformation that a protein undergoes upon complex formation[14]:

$$A_{rel} = \frac{A_{observed}}{A_{monomer}} \qquad (2)$$

Deviations from the idealized $A_{monomer}$ (*i.e.* $A_{rel}$ values > 1) for both monomeric proteins and bound subunits were found to correlate with the magnitude of conformational changes that occur upon binding[14]. Interestingly, the $A_{rel}$ values of a limited set of monomeric proteins were found to correlate with their flexibilities derived from normal mode analysis. Therefore, the correspondence between free-state $A_{rel}$ values and conformational changes could be explained if increasing flexibility in the unbound state were associated with larger binding-induced conformational changes. Roughly speaking, we found that an $A_{rel}$ value of 1.2 tends to correspond to a very large conformational change upon complex formation (> 5 Å root-mean squared deviation between free and bound states). Given that a precision on the order of 50-100 Å² is expected for solvent accessible surface area calculations[24], $A_{rel}$ values can be expected to have a precision of 0.005-0.01 for a typical folded protein with $A_{observed}$ of $10^4$ Å².

Why would solvent accessible surface area be related to intrinsic flexibility? Insight into this can be obtained by considering the relationship with molecular weight shown in Figure 1A. Proteins with high $A_{rel}$ values expose more surface area than expected for typical folded proteins and thus, crucially, bury less surface area intramolecularly within their folds. Since the burial of surface area is the primary driving force that overcomes the huge conformational entropy of the unfolded state[22], proteins with higher $A_{rel}$ values should therefore retain greater conformational entropy and be more flexible. On the other hand, proteins with lower $A_{rel}$ values bury more surface within their folded cores and should thus be more rigid. This concept is illustrated in Figure 1B where the structures and solvent accessibilities of several monomeric proteins are shown along with their $A_{rel}$ values.

This relationship between solvent accessible surface area and protein flexibility becomes even clearer if one assumes a constant relationship between the amount of surface area buried and the energy of folding. While proteins are of course heteropolymers comprised of chemically different amino acids, the mean hydrophobicity of accessible and buried surface area is fairly similar from one protein to another[15]. Moreover, the difference in energetic contribution between buried hydrophobic and hydrophilic surface area should be minimal, assuming that buried polar residues are hydrogen bonded[25–28]. Notably, uniform surface area approaches have been found by some to be much preferable to empirical weighting schemes[29] and have recently shown remarkable agreement with experiments when used to predict protein complex assembly pathways[30,31]. Therefore, by assuming constant surface area, the difference between $A_{observed}$ and $A_{monomer}$ can be directly related to the difference in conformational entropy with respect to the idealized folded state from Equation 1, and could even be converted into approximate energetic terms, for instance by assuming ~0.1 kJ/mol of conformational entropy per additional Å$^2$ exposed compared to the reference state[25]. $A_{rel}$ therefore effectively represents a simple proxy for this difference in conformational entropy, normalized to protein size, explaining why it would be closely related to intrinsic protein flexibility.

### *$A_{rel}$ predicts global measures of intrinsic flexibility*

The above suggests that $A_{rel}$, which functions essentially as a measure of how little surface area a protein buries within its fold, should be related to flexibility. How strong is this relationship, and could it be of practical utility? Here this is investigated in detail through large-scale comparisons of monomer $A_{rel}$ values to multiple independent probes of intrinsic protein flexibility. Although various definitions of flexibility have been used in different contexts, here we are considering intrinsic flexibility in a general sense, as a scalar quantity describing the magnitude of conformational dynamics that a protein will experience isolated in solution. Each flexibility measure used here is very different, being based upon different simulation methods or experimental data, but they are all reflective of intrinsic flexibility in the sense that they quantify the overall extent of protein motions.

A non-redundant set of 6565 monomeric protein crystal structures was first compiled, as well as a subset of 907 high-confidence monomers that have been filtered using much stricter criteria. The high-confidence monomers were used to fit the relationship in Equation 1 (Figure 1A). Figure 1C shows the distribution of $A_{rel}$ values for both sets. A smooth distribution is observed with a peak at 1.0, suggesting that $A_{rel}$ represents a continuous phenomenon with a range of possible values, consistent with the continuum view of protein dynamics. The asymmetric nature of the distribution implies that there are fairly tight constraints on how much surface area a protein can possibly bury intramolecularly (with no observed $A_{rel} <$ 0.8) and looser restrictions on how much surface area a crystallizable monomer can expose (with a few instances of $A_{rel} > 1.5$).

Normal mode analysis provides a fast and simple way to probe the intrinsic flexibility and dynamics of proteins of known structure. Excellent agreement has been obtained between normal mode analysis applied to simple backbone-only models of various proteins and the intrinsic dynamics and conformational changes observed experimentally or in molecular dynamics simulations[32–36]. In this study, the Gaussian network model (GNM)[37] was used for large-scale normal mode analysis. Figure 2A plots the flexibility values calculated with GNM (*i.e.* the mean amplitudes of their normal modes) against $A_{rel}$ for all 6565 monomeric crystal structures. Overall, a strong correspondence is observed (Pearson correlation coefficient $r = 0.78$), confirming the effectiveness of $A_{rel}$ as a proxy for intrinsic flexibility derived from normal mode analysis. Notably, the correlation with the large dataset used here is very similar to what was seen previously with only 60 proteins ($r = 0.76$)[14], using a different method of normal mode analysis (elastic network model[38,39]).

Models of protein structures calculated from NMR data generally contain multiple distinct conformations. While the structural heterogeneity among different ensemble members arises from both the intrinsic dynamics of a protein and of possible uncertainty reflecting an insufficient number of experimental restraints, NMR ensemble models generally provide reasonable, albeit imperfect, representations of the solution dynamics[40–43]. The NMR ensemble models used in this study came from the RECOORD database and have all been recalculated using a uniform protocol[44], thus avoiding some of the variation between models that arises due to methodological differences. Figure 2B plots the root-mean-squared fluctuations (RMSF) for 454 non-redundant NMR models against their $A_{rel}$ values. Overall, there is a strong correlation between $A_{rel}$ and RMSF ($r = 0.82$), demonstrating that $A_{rel}$ is highly reflective of the fluctuations within an NMR model.

Molecular dynamics simulations provide another way to characterize the intrinsic flexibility and dynamics of proteins in detail. This study utilized a large set of 10 ns trajectories from the MoDEL database, which contains pre-calculated parameters describing the global flexibility of each protein[45]. Figure 2C compares $A_{rel}$ to the Lindemann index calculated from the molecular dynamics trajectories of 491 non-redundant monomeric proteins. This parameter provides a measure of atomic-level disorder, and can thus be used a descriptor of the liquid-like or solid-like character of a protein[46]. The correlation is strong ($r = 0.77$) confirming that $A_{rel}$ shows major overlap with flexibility derived from molecular dynamics simulations. Likewise, an alternate measure of flexibility available in MoDEL, the mean variance of C$\alpha$ atoms, shows a similar correlation of 0.73 with $A_{rel}$ (Figure S1).

The Random Coil Index (RCI) is based upon NMR chemical shifts and provides a local measure of backbone flexibility[47]. Figure 2D plots the global RCI values, averaged over all residues, against $A_{rel}$ values calculated from 185 non-redundant NMR models. The correlation is good ($r = 0.71$), although slightly lower than the other examined flexibility measures. Importantly, RCI is directly calculated from experimental chemical shifts, which were not used in the NMR model calculations, and is thus independent of the structural models used to calculate $A_{rel}$.

Together, the above results demonstrate that the overall flexibility of monomeric proteins is strongly determined by the total amount of surface area buried within their folds, and thus the simple-to-calculate $A_{rel}$ parameter is highly predictive of various flexibility measures. Table S1 shows how those different measures correlate with each other. For example, while the Lindemann index shows a correlation of 0.77 with $A_{rel}$ values, it has markedly lower correlations with RMSF (0.70), RCI (0.57) and GNM (0.68). Strikingly, the $A_{rel}$ values correlate with all four measures of intrinsic flexibility as well as, or better than, those measures do with each other. Furthermore, it was also found that the choice of monomer set used for fitting Equation 1 had little effect on the predictive ability of $A_{rel}$ (Table S2).

## *Flexibility depends on protein length and number of domains*

$A_{rel}$, by definition, is essentially normalized to protein length ($r$ = 0.999 between number of residues and molecular weight in the full dataset), and reflects the flexibility with respect to an idealized state expected for a protein of a given molecular weight. However, there does appear to be some length dependence to protein flexibility, as there is a slight correlation between $A_{rel}$ and chain length ($r$ = 0.17). One possibility is that this is related to the number of protein domains, as the presence of multiple domains can facilitate considerable inter-domain motions[5,48]. Indeed, the correlation between $A_{rel}$ and number of domains is much stronger ($r$ = 0.31), while the correlation with chain length largely disappears if one considers only single-domain proteins ($r$ = 0.07). Figure 3 demonstrates the strong tendency for mean $A_{rel}$ values to increase with an increasing number of domains.

Could the motions facilitated by the presence of multiple domains be a primary determinant of the relationship observed between $A_{rel}$ and other measures of flexibility? Table S3 shows that this is not the case by breaking down the analyses by single- and two-domain proteins, and showing that the strong correlations are still preserved. Thus $A_{rel}$ clearly captures much more of the broad spectrum of protein flexibility than just inter-domain motions.

The influence of multidomain proteins on the fit in Figure 1A used to derive Equation 1 was also considered. If multidomain proteins are excluded, Equation 1 changes to $5.82M^{0.74}$. This suggests that, unexpectedly, the approximation of proteins adopting relatively constant shapes becomes slightly more accurate when considering only single-domain proteins, as the exponent is closer to 2/3. However, most importantly for our purposes, the correlations with different measures of intrinsic flexibility change very little if this form of the equation is used to calculate $A_{rel}$ (Table S2).

## *Flexible proteins are characterized by unique sequence and secondary structure properties*

To what extent is the propensity for protein flexibility reflected in amino acid composition? To assess this, the Pearson correlation coefficient (*r*) was calculated between the fractional content of each of the 20 standard amino acids and the $A_{rel}$ values of monomeric crystal structures (Figure 4A). Thus amino acids that tend to be associated with higher-$A_{rel}$, more flexible proteins will have higher *r* values, while those associated with more rigid proteins will have negative *r*. Each amino acid is coloured by its type (hydrophobic, charged, polar and glycine), which allows some interesting trends to be immediately noted. Charged amino acids, with the exception of aspartate, have a strong association with increased flexibility. On the other hand, polar amino acids are generally associated with lower flexibility. Interestingly, and somewhat surprisingly, hydrophobic amino acids tend to be intermediate between polar and charged residues. Finally, glycine shows the strongest negative correlation with $A_{rel}$. Very similar sequence trends are observed if alternate measures of flexibility are considered (Figure S2A).

Some correspondence between the sequence determinants of intrinsic disorder and intrinsic flexibility as measured by $A_{rel}$ might be expected. For instance, protein complex subunits predicted to be disordered have been found to have much higher $A_{rel}$ values than those predicted to fold[14,49]. Contact density, which is closely related to buried surface area, has also been identified as an important parameter for predicting intrinsically disordered regions[50–52]. Furthermore, regions of proteins predicted to be disordered yet observed in crystal structures tend to undergo larger conformational changes[53]. Therefore, it is noteworthy that the association between $A_{rel}$ and charged residues is analogous to the previous observations that net charge is the primary determinant of expandedness in intrinsically disordered proteins[54–56]. Beyond this, however, there appears to be little further similarity with the known sequence determinants of disorder[7,8]. For example, glycine is strongly associated with intrinsic disorder yet here inversely correlates with $A_{rel}$. Moreover, leucine is relatively rare in disordered proteins, yet here shows one of the strongest correlations with increased flexibility. Table S4 shows there are only weak correlations between the $A_{rel}$ values of monomeric proteins and multiple sequence-based disorder predictions, suggesting that, for the most part, intrinsic disorder and monomer flexibility as manifested by $A_{rel}$ are reflective of different regions of the protein dynamics continuum.

In contrast to intrinsic disorder, there does appear to be a clear association between $A_{rel}$ and the secondary structure propensities of different amino acids. In particular, glutamate, leucine and lysine have strong α-helical propensities, while glycine, tyrosine and asparagine are helix destabilizing[62]. Therefore, given this apparent correspondence between flexibility and secondary structure propensities, the $A_{rel}$ values of monomer crystal structures from different SCOP classes[58] were compared (Figure 4B). Consistent with the sequence trend, this analysis reveals that all-α proteins are the most flexible (mean $A_{rel}$ = 1.050) and all-β proteins the most rigid (mean $A_{rel}$ = 0.984, $P < 2.2 \times 10^{-16}$, Wilcoxon test). The mixed classes (α+β and α/β) have $A_{rel}$ values intermediate to α and β, although α/β and β are nearly equal. This tendency for α proteins to be more flexible than β proteins maintained when alternate measures of flexibility are considered instead of $A_{rel}$ (Figure S2B). Furthermore, the sequence trends and correlations between $A_{rel}$ and different measures of flexibility are preserved when split by structural class, demonstrating that they are largely independent of secondary structure (Figure S3 and Table S5).

What is the origin of this difference in flexibility between α and β proteins? One explanation is that β-strands are more often associated with changes in the direction of the polypeptide chain. Thus one can easily imagine why β proteins would tend to form more compact, low-$A_{rel}$ structures that bury more surface area within their folds. In contrast, α-helices only require the chain to go in a single direction, so more extended, high-$A_{rel}$ structures can be facilitated by the presence of helical structure. It is also interesting that many of the most compact, low-$A_{rel}$ structures contain a mixture of both α and β structure, as for example seen in the TIM barrel β-mannanase shown in Figure 1B (PDB ID: 1BQC); in these cases the combination of α and β structure may facilitate their highly pseudosymmetrical folds. Interestingly, α/β proteins were recently shown to be the most compact structural class and to exhibit the slowest folding rates, which was attributed to the fact that they were able to adopt the most spherical structures[59–61]. This helps to explain how these proteins are able to so efficiently bury surface area and adopt overall rigid conformations. In light of these observations, it is interesting to note the recent finding that proteins with β structure tend to have slightly different $A_s$ vs. molecular weight relationships (*i.e.* Equation 1) than all-α proteins[63], which could imply that larger β proteins tend to be relatively more flexible than larger α proteins, in comparison with smaller proteins.

These results could also possibly be related to the strong relationship between low $A_{rel}$ values and polar residues, if the requirement for these side chains to form intramolecular hydrogen bonds is associated with changes in backbone direction, *e.g.* for stabilizing turns or long-range β-strand contacts[57]. Similarly, the strong association between glycine and rigidity could be related to its favourability for forming type-II β-turns. Thus, while the vast Ramachandran space available to glycine is often associated with local flexibility, the results here suggest that this also gives it the ability to facilitate globally rigid structures that can effectively bury surface area and stabilize their folds.

## *Quaternary structure errors are associated with high apparent flexibility*

Careful examination of the unit cells of monomeric crystal structures with high $A_{rel}$ values suggested that some of these were actually homo-oligomeric, with the monomeric biological unit likely being assigned in error. For example, the structure of TrmD (PDB ID: 1P9P) is shown in Figure 5A. This structure has an author-assigned monomeric biological unit with a high $A_{rel}$ value (1.29), yet it has been manually annotated as a dimer in the PiQSi database of manually curated quaternary structure (QS)[64] and is also predicted to be dimer by PISA[65].

To investigate the relationship between apparent flexibility and the propensity for QS misassignments, correctly and incorrectly assigned monomers were identified from PiQSi. On average, confirmed monomers tend have lower $A_{rel}$ values (1.01) than those assigned in error (1.08, $P = 2 \times 10^{-6}$, Wilcoxon test), thus demonstrating that incorrectly assigned monomers are associated with greater apparent flexibility. Figure 5B shows the frequency of QS errors for proteins grouped by $A_{rel}$ values. Notably, there is a strong tendency for the frequency of QS errors to increase with higher $A_{rel}$. Thus while $A_{rel}$ alone would not be able to absolutely identify QS errors, this plot can be used to roughly assess the likelihood that an apparently monomeric protein with a given $A_{rel}$ value is the result of a QS misassignment. This could potentially be of considerable use in QS-prediction algorithms, which incorporate many factors.

## *Intrinsic flexibility is a major determinant of crystal structure resolution*

Decades of experience have shown that flexible proteins are generally difficult to crystallize. Therefore, it is natural to ask whether there might be a relationship between the intrinsic flexibility of a protein and the resolution of its crystal structure. Figure 6A plots the mean $A_{rel}$ values for monomeric crystal structures grouped by resolution. Interestingly, there is a marked tendency for proteins with lower resolution crystal structures to be more flexible. For instance, very high-resolution monomers (< 1 Å) have a mean $A_{rel}$ of 0.97, compared to 1.19 for those of low resolution (≥ 3.5 Å) ($P = 4 \times 10^{-8}$, Wilcoxon test). This relationship is preserved even when only crystal structures with no disordered residues are considered and is also confirmed with different measures of flexibility (Figure S4).

An interesting consequence of this result is that, as protein crystallography has experienced methodological improvements, the ability to solve lower resolution crystal structures and thus probe more flexible regions of protein conformational space has improved. Figure 6B shows the mean $A_{rel}$ values of monomeric crystal structures solved over time. A clear tendency is observed for more recently determined crystal structures to represent more flexible proteins. Therefore, as experimental and computational methods for modelling lower resolution crystal structures continue to improve[66], the available coverage of the protein dynamics continuum will continue to broaden.

## *Residue-specific $A_{rel}$ reflects local flexibility*

Since $A_{rel}$ is based upon the total solvent accessible surface area, it therefore provides no information on local protein flexibility. However, we can easily employ a residue-specific local $A_{rel}$ measure, defining it as the ratio of the observed solvent accessible surface area for a residue to the expected unfolded-state value for that amino acid type[15]. Similar approaches have been used in the past, for instance in defining buried and accessible residues[67]. In addition, local solvent accessibility is known to be closely related to B factors from crystal structures[68–70] and order parameters from NMR relaxation experiments[71]. Therefore, to complement the above global analyses, local $A_{rel}$ values have been compared to residue-specific flexibility values derived from normal mode analysis, NMR ensemble models and chemical shifts.

The distributions of Pearson correlation coefficients calculated for individual proteins between local $A_{rel}$ values and local flexibilities calculated from different methods are shown in Figure 7A-C. Overall, mean correlations of 0.74 for GNM, 0.70 for RMSF and 0.61 for RCI are observed. Thus local $A_{rel}$ provides fairly reasonable predictions of local flexibility for all of these methods. The origin of this relationship is the same as discussed earlier for global flexibility: proteins with lower local $A_{rel}$ values bury more surface area, making more extensive intramolecular contacts, and are thus more motionally restricted. In effect, local $A_{rel}$ here is functioning similarly to the simple contact models showing that flexibility is primarily determined by local contact density, thus allowing prediction of B factors, NMR order parameters[72,73] or intrinsic disorder[51,52].

## Conclusion

Although it has long been clear that protein flexibility is important for function, characterizing this flexibility can be difficult. Here it is shown that $A_{rel}$, which functions essentially as a proxy for how much surface area a protein buries within its fold, correlates remarkably well with different measures of intrinsic protein flexibility. This allows the easy assessment of protein flexibility in a quantitative manner from the large number of protein structures now available, and has revealed new insight into the relationships between protein flexibility, sequence and structure. Many more topics of inquiry remain open for future exploration, relating to diverse aspects of protein structure, function, sequence and evolution.

The major advantage of $A_{rel}$ as a probe of protein flexibility is its simplicity. It can be quickly and easily calculated from any protein structure. Furthermore, its correlation with intrinsic flexibility arises directly from the fundamental energetics of protein folding, so its utility is not merely empirical. While treating surface area uniformly works remarkably well for many purposes, it is possible that a model considering the specific properties of a protein's surface could provide a better probe of flexibility. However, the fact that $A_{rel}$ correlates as good or better with different measures of intrinsic flexibility than they do with each other suggests that any room for improvement with a more complex model should be minimal.

Another potential issue relevant to this study relates to crystallographic disorder, which results in residues missing from the crystal structures. Since this study dealt primarily with structure-based measures of flexibility here, the effect of ignoring disordered residues should be small. However, in principle, disordered residues could be dealt with, either by explicit modelling[74], or by simply assigning them statistical accessibilities, with the assumption that they do not form any non-local contacts. Importantly, all trends in this study are retained when considering only high-confidence monomers with no disordered residues.

Previously, $A_{rel}$ values of both monomeric proteins and bound subunits were used to predict protein conformational changes upon binding[14]. Analysis of the free/bound pairs shows a very strong correlation between the $A_{rel}$ values of monomeric proteins and their $A_{rel}$ values in the bound state. This suggests that, in addition to their utility for predicting conformational changes upon binding, the $A_{rel}$ values of bound subunits are predictive of the flexibility of proteins in their free states, thus facilitating large-scale analyses of how intrinsic flexibility relates to protein complex assembly from the structures of protein complexes.

Finally, it is interesting to note that the unique amino acid properties associated with intrinsic flexibility appear to be distinct from intrinsic disorder. Given the success of intrinsic disorder predictors, it is tempting to speculate that a sequence-based predictor of flexibility could be developed using $A_{rel}$ values as a training set. Previously B factors have been used in a similar manner to train sequence-based flexibility predictors[75,76], although the amino acids associated with high B factors are quite different than the sequence determinants of flexibility observed here, likely due to the differences between global and local flexibility. It is possible that the topological complexity of protein folds might inhibit a sequence-based predictor of folded protein flexibility, as compared to intrinsically disordered proteins where the amino acid composition can sometimes be more important than the specific linear sequence[77]. However, the availability of a sequence-based predictor would facilitate genome-scale analyses of the relationships between protein flexibility, function and evolution.

# Methods

## *Monomer datasets*

All monomeric crystal structure biological units containing at least 30 residues were taken from Protein Data Bank on 2012-08-08, excluding backbone-only models. The set of high-confidence monomers (used for fitting the relationship in Figure 1A) included only monomers with SCOP 1.75 domain assignments[58] in order to specifically exclude structures in the classes "membrane and cell surface proteins and peptides", "small proteins", "coiled coil proteins", "low resolution protein structures", "peptides" and "designed proteins". In addition, only proteins which had a single chain in the asymmetric unit and which were predicted to be monomeric by PISA[65] were considered. Furthermore, structures which contained >= 1% non-protein atoms were excluded. Finally, proteins in which any non-terminal residues were missing from the crystal structure (*i.e.* disordered) were excluded.

The full crystal-structure dataset used for most of this paper also included only proteins that were predicted to be monomeric by PISA, and excluded structures which contained >= 10% non-protein atoms. In addition, any structures containing a large number of missing, non-terminal residues (containing any stretches of > 20 missing residues, or > 50 missing residues in total) were excluded. For the dataset used in the QS-assignment analysis, the same criteria were used, except for the condition that proteins be PISA-predicted monomers.

The NMR ensemble models were taken from the RECOORD database and have all been recalculated using a uniform protocol and refined in water[44].

For all datasets, sequence redundancy filtering was performed at a level of 90% identity, after the above criteria were applied. This left 907 non-redundant proteins in the high-confidence set, 6565 in the full set, 491 with MoDEL parameters, 454 in the NMR set and 267 in the set with PiQSi QS assignments. All protein structures and relevant parameters related to these analyses are provided in Table S7.

## *Structural and flexibility calculations*

### *Solvent accessible surface area*

Solvent accessible surface area was calculated from each crystal structure and each NMR model ensemble member using AREAIMOL[78]. For calculating $A_{rel}$, the solvent accessible surface area was averaged over all NMR ensemble members. For the local $A_{rel}$ calculations, the unfolded-state values of Miller *et al.* were used[15], and values were averaged over a seven-residue window. All correlations with $A_{rel}$ were calculated using the log of the flexibility parameters, consistent with what was done previously for conformational changes[14] since this tends to give much more linear relationships.

*Normal mode analysis*

The Gaussian network model[37] was used with default parameters and considering only backbone C$\alpha$ atoms. For each protein, $n$ normal modes are calculated with GNM for a protein with $n + 1$ residues. The flexibility of each protein was calculated as the average of the inverse eigenvalue (*i.e.* frequency, $\omega_i$) of each normal mode (Equation 3). This value therefore represents the mean amplitude of a protein's normal modes.

$$GNM(global) = \frac{\sum_{i}^{n} 1/w_i}{n} \quad (3)$$

The residue-specific flexibility for residue $j$ is given in Equation 4, where $a_{ij}$ is the displacement of residue $j$ under mode $i$.

$$GNM(local)_j = \sum_{i}^{n} \frac{a_{ij}}{w_i} \quad (4)$$

*NMR models*

For each NMR ensemble model, the residue-averaged RMSF was calculated according to Equation 5, where $d_{ij}$ is the distance between each backbone C$\alpha$ atom $i$ from conformer $j$ and the ensemble-averaged position of that atom, $m$ is the total number of conformers in the ensemble, and $n$ is the total number of residues in the protein.

$$RMSF = \frac{\sum_{i}^{n} \sqrt{\frac{\sum_{j}^{m} d_{ij}^2}{m}}}{n} \qquad (5)$$

*Molecular dynamics simulations*

All parameters were taken directly from the MoDEL database and were calculated from 10 ns trajectories using AMBER8 or AMBER9 with parm99 force field and TIP3P water model[45]. Because trajectories in MoDEL were only calculated for a limited subset of all available monomeric proteins, a separate 90% sequence identity redundancy filtering was performed with these proteins.

*NMR chemical shifts*

The BioMagResBank[79] was searched for sets of chemical shifts that corresponded to our non-redundant NMR models. In total, chemical shifts were available for 185 of these proteins. The Random Coil Index (RCI) values were then calculated from these chemical shifts using the RCI web server[47]. For the global RCI measure, the residue-specific values were averaged over all positions.

## Acknowledgements

I thank Tim Meyer from providing data from MoDEL and Sarah Teichmann, Julie Forman-Kay, Tina Perica and A.J Venkatakrishnan for valuable comments on the manuscript. This work was supported by a Long-term Fellowship from the Human Frontier Science Program.

# Figures

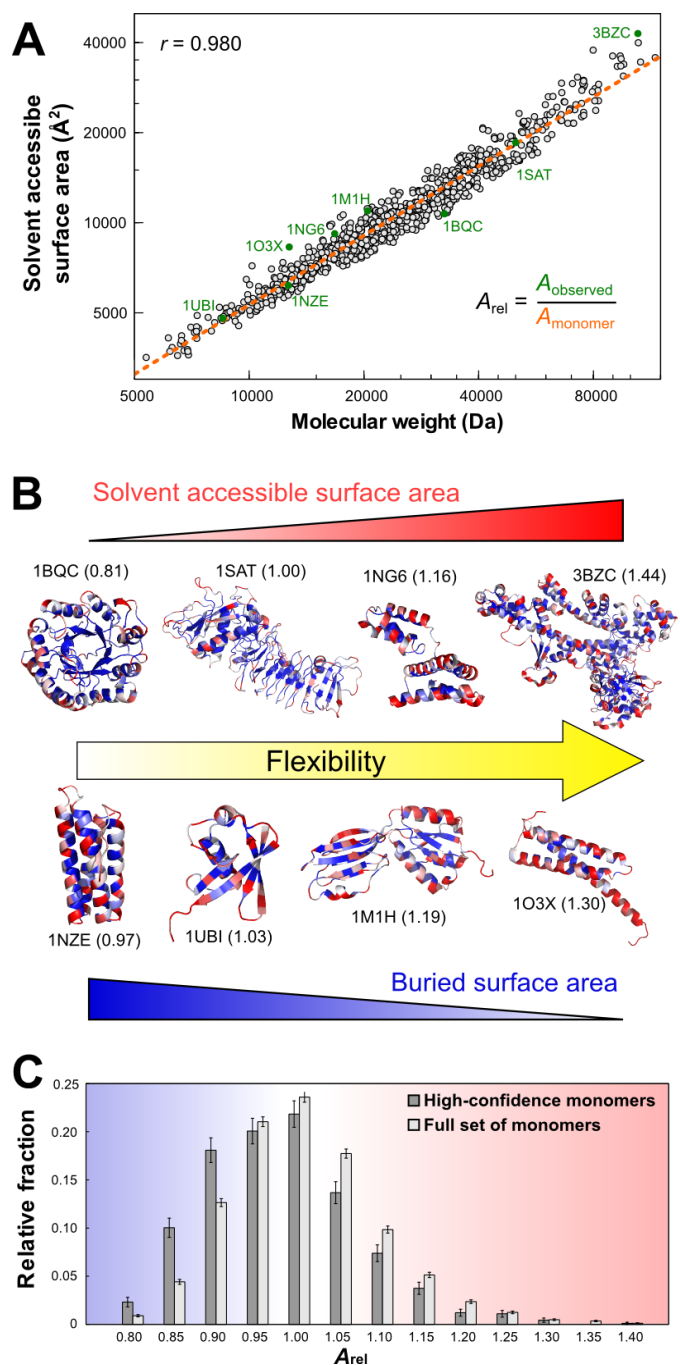

**Figure 1. Relative solvent accessible surface area ($A_{rel}$) values of monomeric proteins.** (A) Comparison of molecular weight and solvent accessible surface area values calculated from 907 non-redundant high-confidence folded, monomeric crystal structures. (B) Crystal structures of monomers of varying $A_{rel}$ (given in brackets), with buried residues coloured blue and solvent exposed residues coloured red. These proteins are highlighted in green in panel (A). (C) Distributions of $A_{rel}$ values for the high-confidence monomers and the full set of 6565 non-redundant monomeric crystal structures. Error bars represent the standard error of the mean.

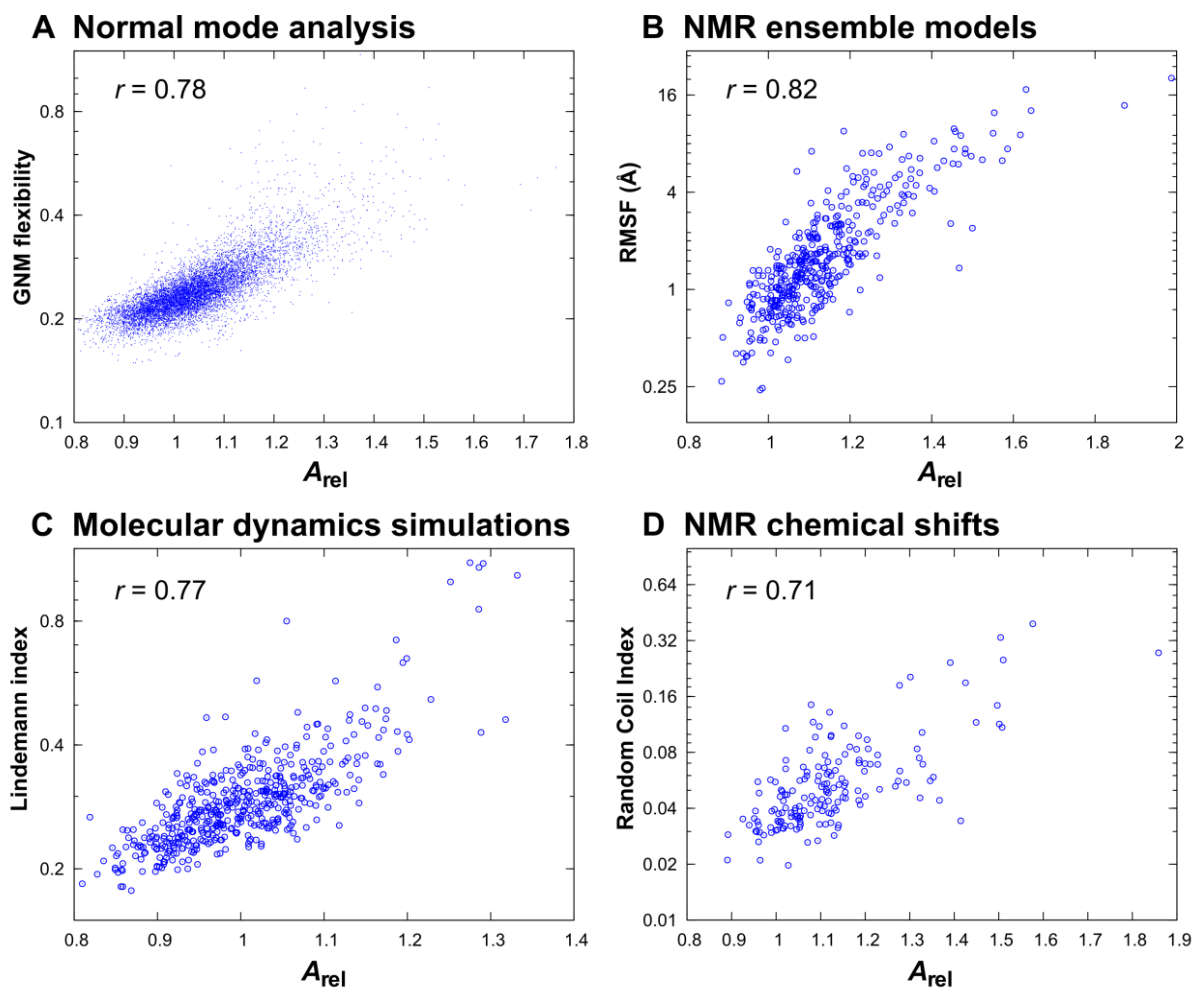

**Figure 2. Comparison of monomer $A_{rel}$ values to independent measures of intrinsic flexibility.** (A) Flexibility of 6565 crystal structures calculated from Gaussian network model (GNM) normal mode analysis. (B) Root-mean-squared fluctuations (RMSF) between conformers calculated from 454 NMR models. (C) Lindemann index calculated from 491 molecular dynamics simulations in MoDEL. (D) Mean Random Coil Index (RCI) values calculated from NMR chemical shifts measured for 185 proteins. Correlations with $A_{rel}$ are calculated with the log of the flexibility parameters, as plotted.

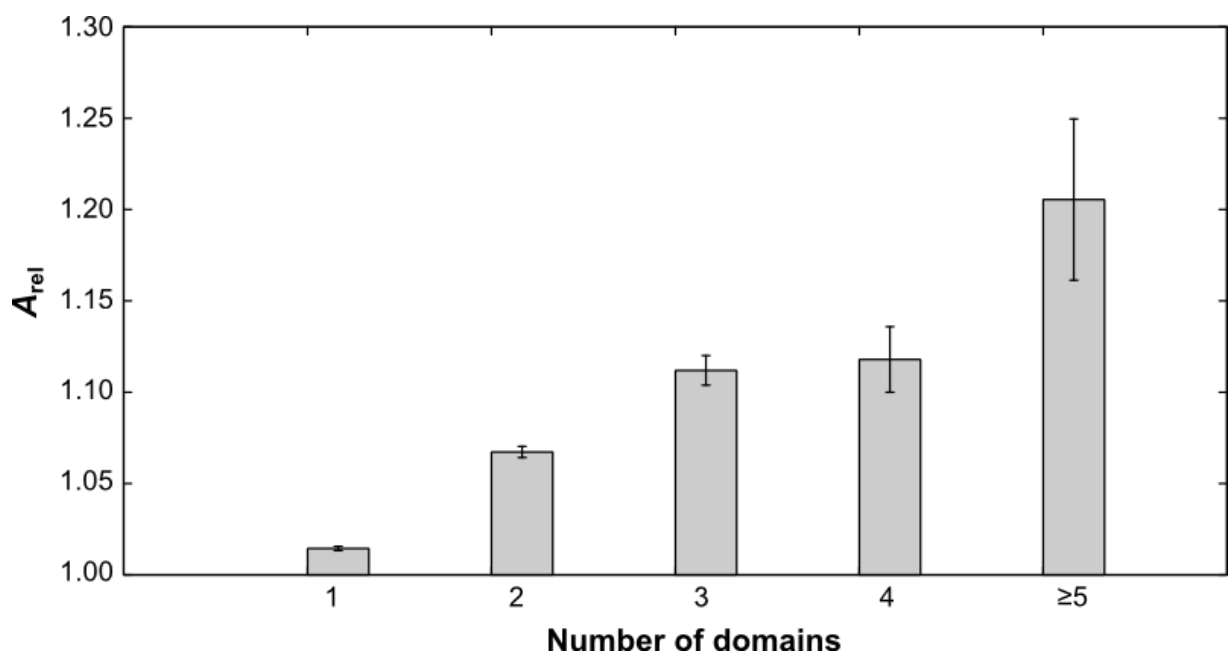

**Figure 3. Association of single and multidomain proteins with intrinsic flexibility.** Comparison of mean $A_{rel}$ values from the full set of monomeric crystal structures grouped by their total number of domains. SUPERFAMILY domain assignments were used since manual SCOP assignments are not available for most proteins in the set. Error bars represent the standard error of the mean.

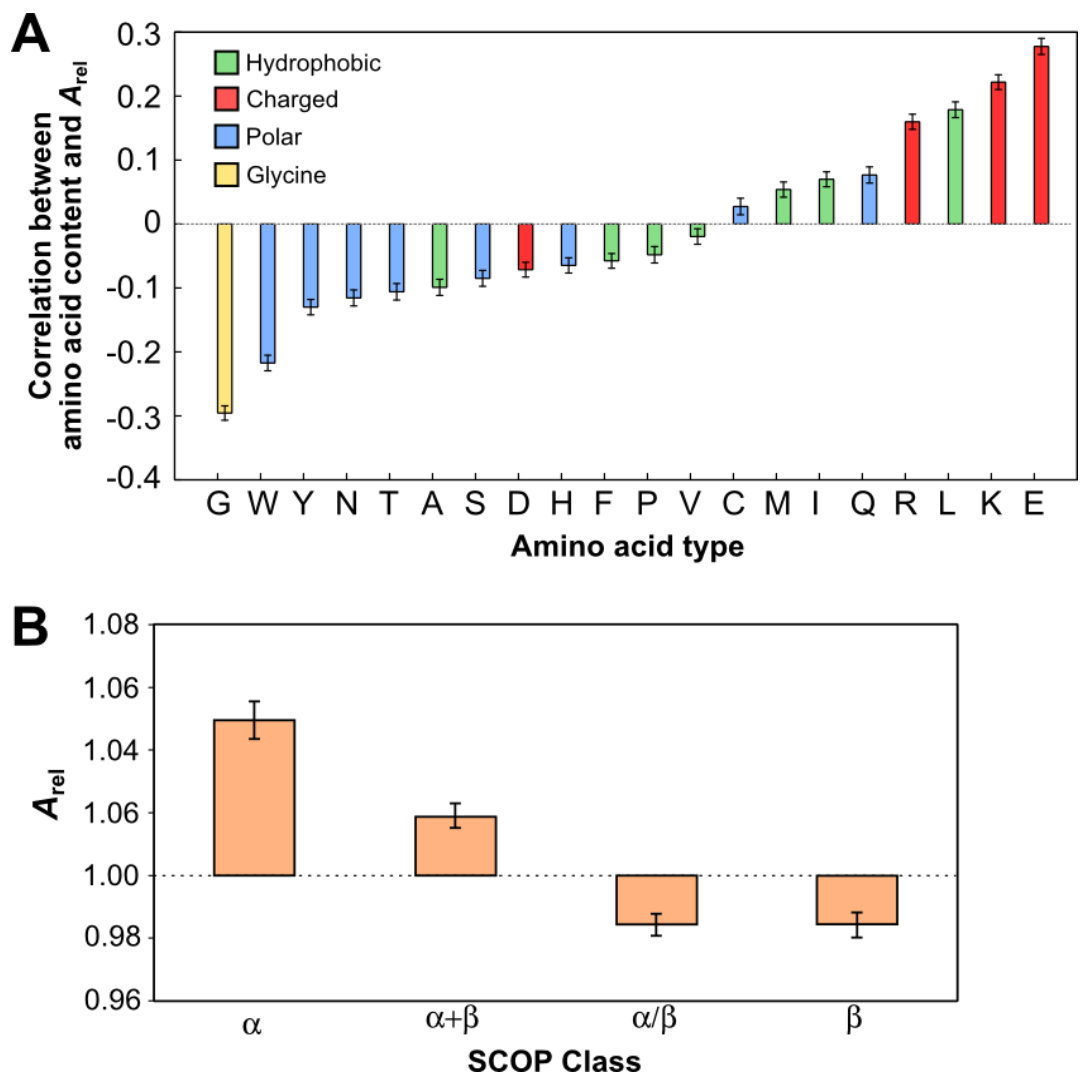

**Figure 4. Sequence and secondary structure propensities associated with intrinsic protein flexibility.** (A) Pearson correlation coefficients (*r*) between the fractional amino acid content and $A_{rel}$ values of 6565 monomeric crystal structures. Error bars represent the standard error of the mean and were calculated from 1000 bootstrapping replicates in the same way as previously described [55]. (B) Comparison of mean $A_{rel}$ values from monomeric crystal structures grouped by SCOP class: all-α, all-β, α+β (segregated α and β regions) and α/β (β-α-β units). Only proteins with a single SCOP domain assignment were considered. All *P*-values were calculated with the Wilcoxon rank-sum test. Error bar represent the standard error of the mean.

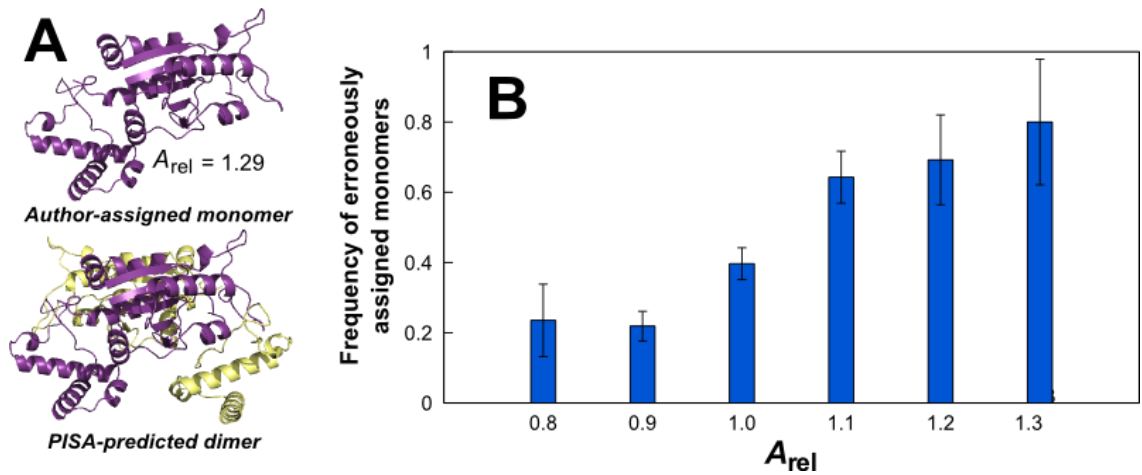

**Figure 5. The relationship between intrinsic flexibility and quaternary structure assignment errors.** (A) Crystal structure of the protein TrmD (PDB ID: 1P9P) which has an author-assigned monomeric biological unit (top), but which is manually annotated in PiQSi and predicted by PISA to be a homodimer (bottom). (B) Frequency of erroneous monomers for proteins with different $A_{rel}$ values. Error bars represent the standard error of the mean.

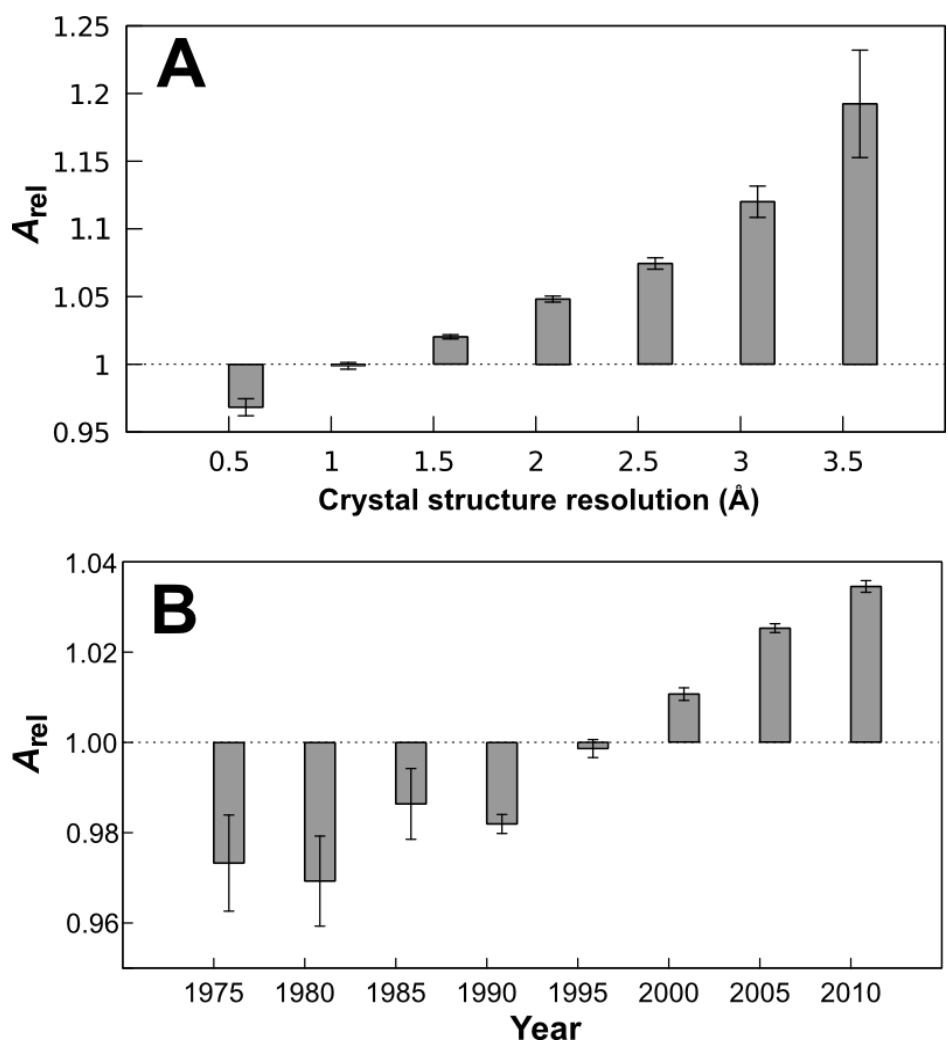

**Figure 6. The relationship between intrinsic protein flexibility, crystal structure resolution and the year of structure determination.** (A) Comparison of mean $A_{rel}$ values from 6565 monomeric crystal structures of different resolutions. (B) Comparison of crystal structure $A_{rel}$ values grouped by the year of structure determination. Error bars represent the standard error of the mean.

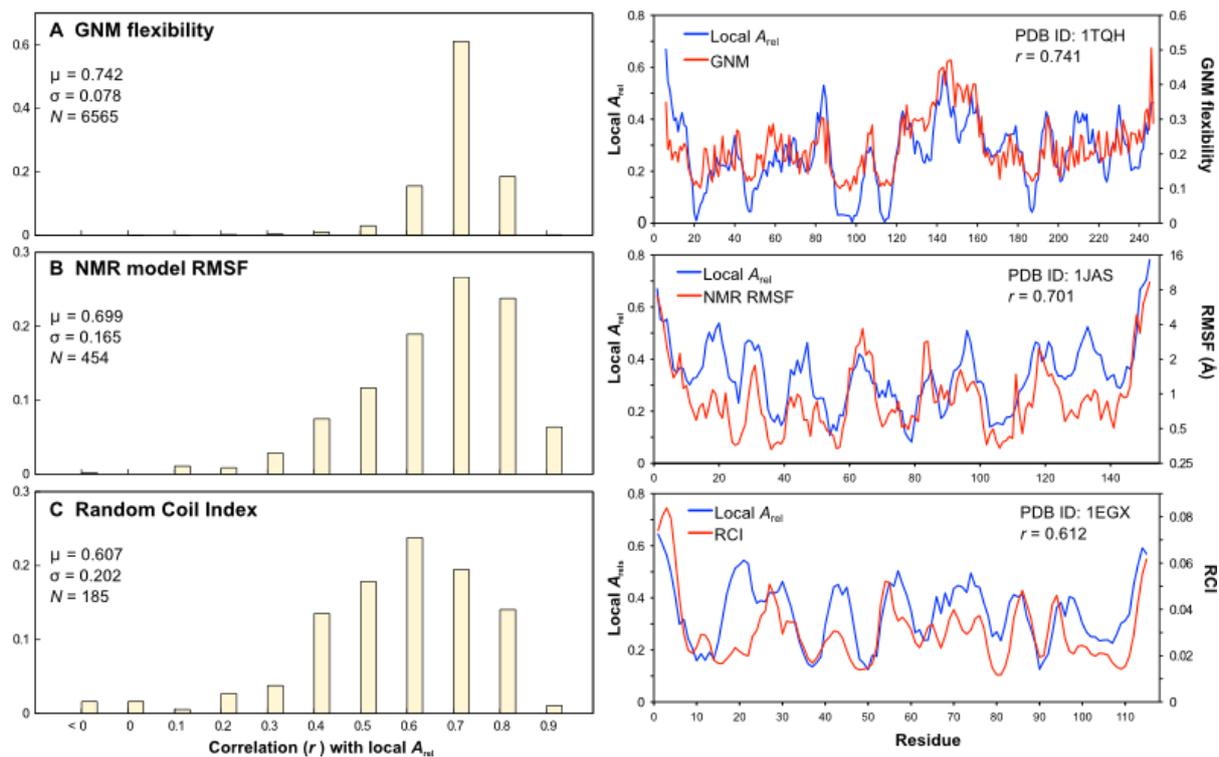

**Figure 7. Comparison of local $A_{rel}$ values to residue-specific measures of protein flexibility.** The distribution between Pearson correlation coefficients (*r*) for individual proteins calculated between local $A_{rel}$ values and (A) flexibility values derived from GNM normal mode analysis, (B) local RMSF values from NMR ensemble models, and (C) Random Coil Index values. On the right, a typical example for each measure is compared to local $A_{rel}$ values. Correlations between different measures of local flexibility are provided in Table S6.

# Supplementary Material

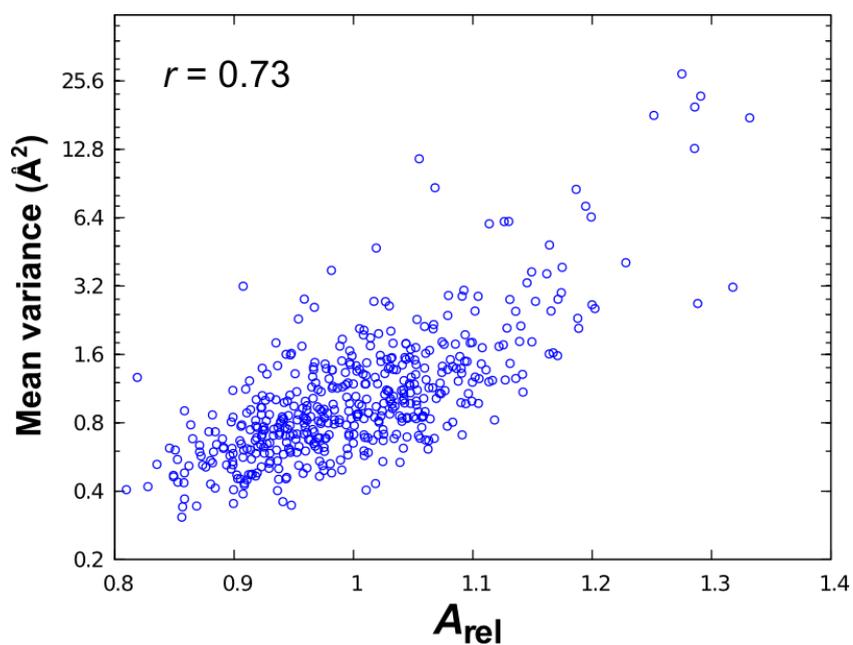

**Figure S1.** Comparison of $A_{rel}$ values from 489 non-redundant monomeric crystal structures with the mean variance of C$\alpha$ atoms calculated from MoDEL molecular dynamics simulations. Note that two of the proteins used for the Lindemann index comparison do not have variance values in MoDEL, thus explaining the slightly smaller dataset.

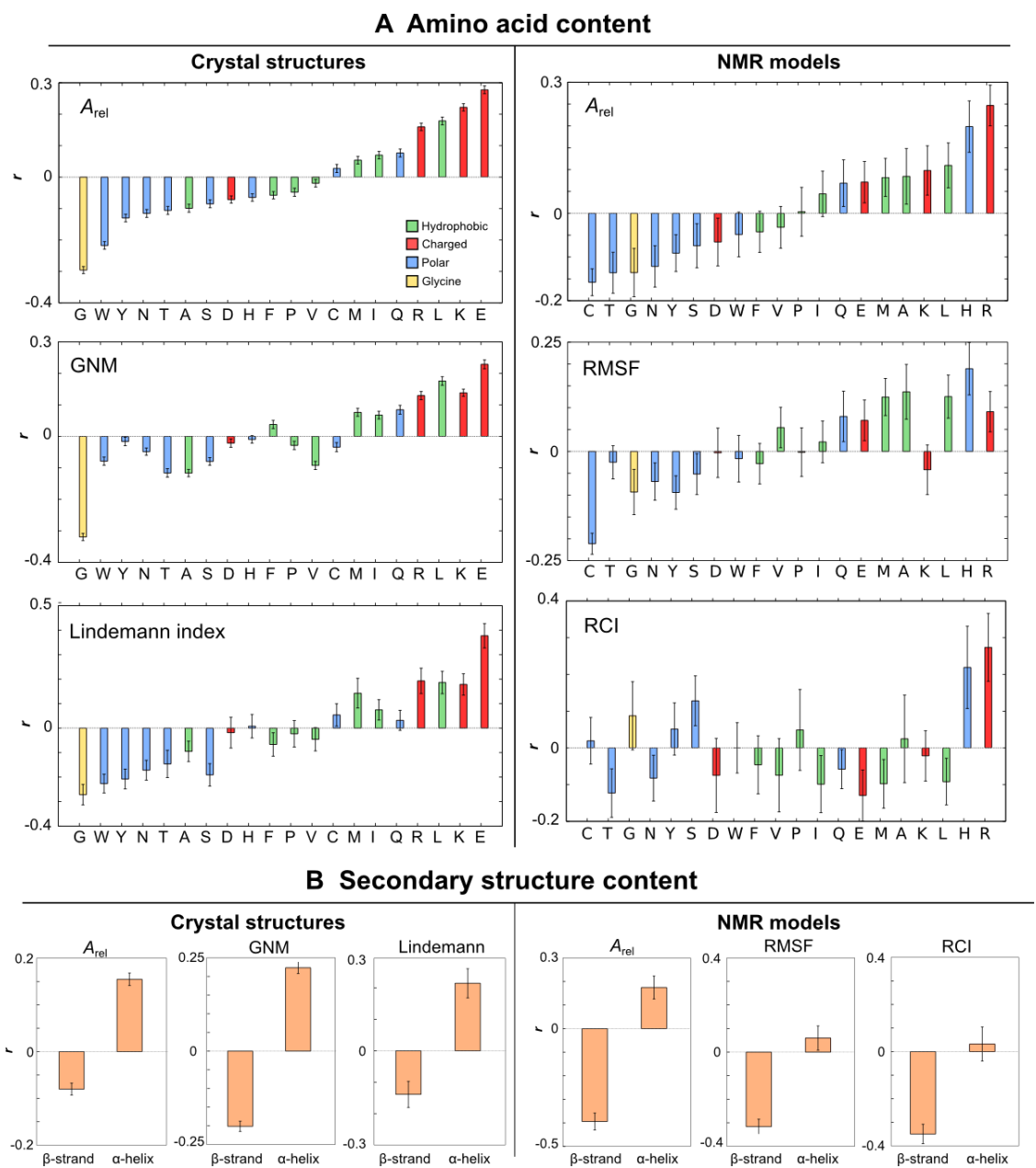

**Figure S2.** Correlations of different flexibility measures with amino acid and secondary structure content. (A) Exactly as in Figure 5A, except that different flexibility measures are used instead of $A_{rel}$. Importantly, the trend of glycine < polar < hydrophobic < charged is largely preserved across different flexibility measures. There are, however, notable differences between the crystal structures and NMR models. In particular, histidine is associated with increased flexibility, probably reflecting the fact that NMR experiments are commonly performed in slightly acidic buffers in which the histidine side chain would be charged. Although RCI shows the largest deviations from $A_{rel}$, note the large size of the error bars due to the much smaller dataset. (B) Similar to (A), except that fractional α-helix and β-strand content is used instead of amino acid content. Crucially, α-helices are associated with flexibility while β-strands are associated with more rigid proteins using all measures of protein flexibility. Secondary structure content was calculated from protein structures with STRIDE[1]. Error bars represent standard error from 1000 bootstrapping replicates, as in Figure 4A.

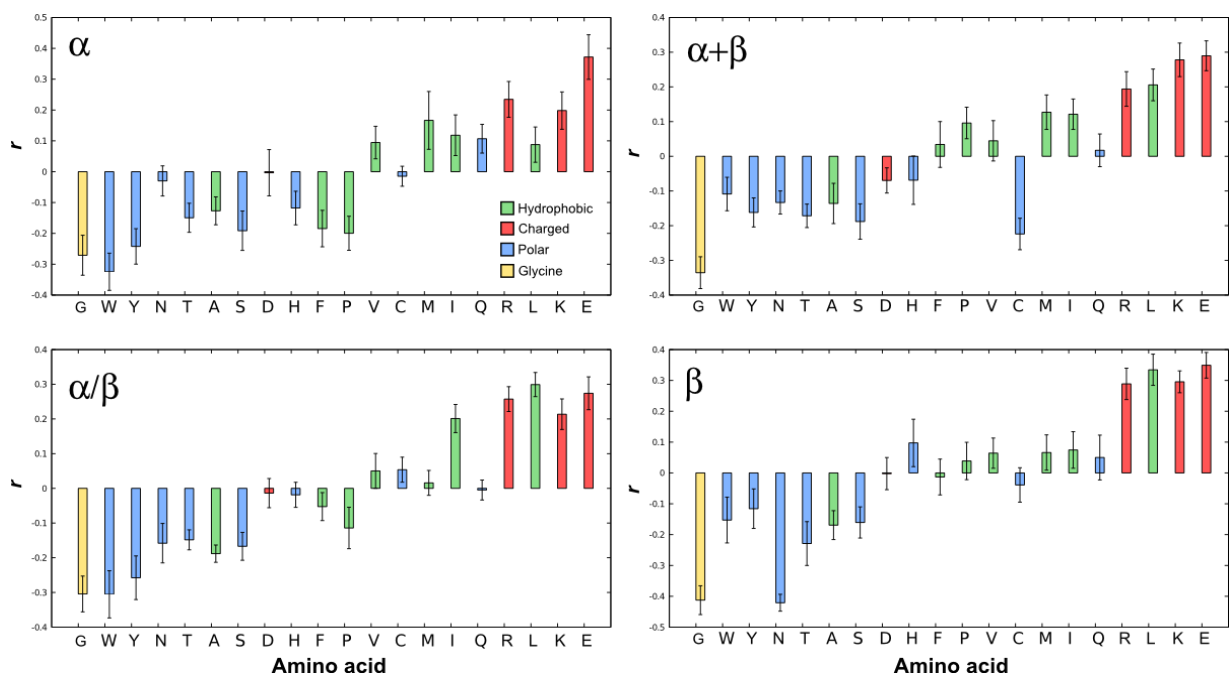

**Figure S3.** Correlations between fractional amino acid content and $A_{rel}$ values for monomeric crystal structures from different SCOP classes. Note that although there are some differences, the main sequence determinants of flexibility are largely preserved across different secondary structural classes.

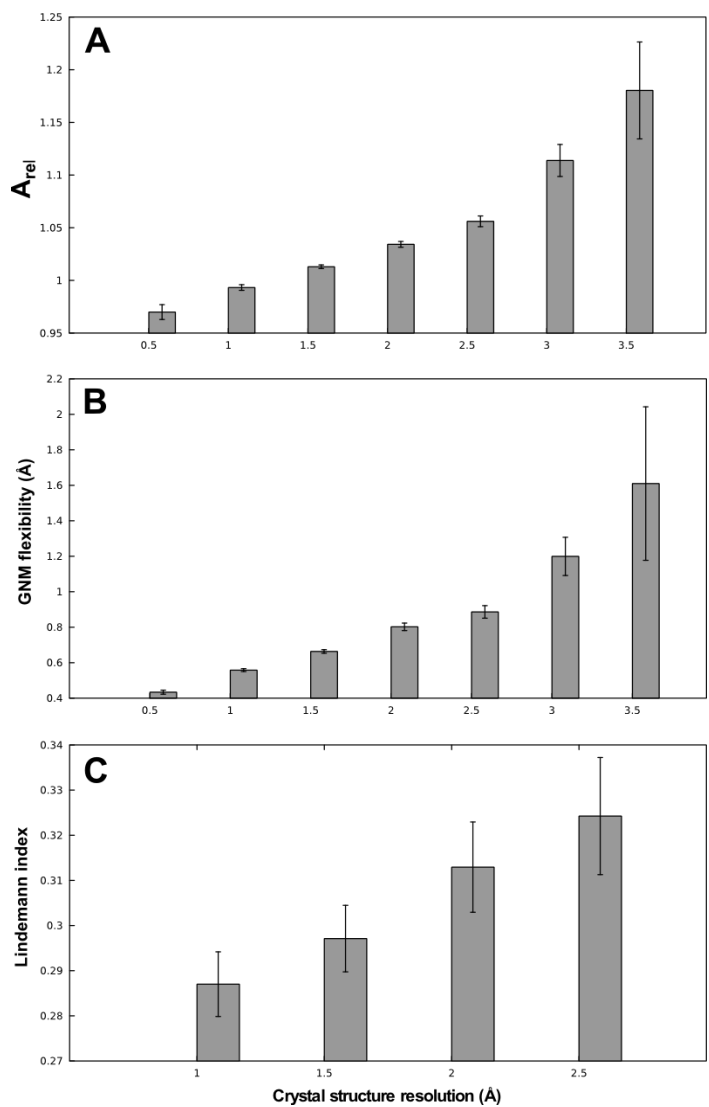

**Figure S4.** Comparison of different flexibility measures (A, $A_{rel}$; B, GNM; C, Lindemann index) to crystal-structure resolution for monomeric structures with no missing residues. The dataset used for this analysis was filtered with the same criteria as the full dataset, except that no non-terminal disordered residues were allowed. Only bins containing at least 10 structures are shown, which accounts for the fewer bins in the smaller molecular dynamics dataset.

**Table S1.** Pearson correlation coefficients between independent measures of global flexibility.

|  | GNM | RMSF | Lindemann | RCI |
|---|---|---|---|---|
| $A_{rel}$ | 0.78 | 0.82 | 0.77 | 0.71 |
| **GNM** | - | 0.78 | 0.69 | 0.56 |
| **RMSF** | - | - | 0.70 | 0.55 |
| **Lindemann** | - | - | - | 0.57 |

Correlations are all calculated using the log of the flexibility parameters, except $A_{rel}$, for consistency with the results presented in the main text.

**Table S2.** Pearson correlation coefficients between $A_{rel}$ values and different measures of intrinsic flexibility, as in Figure 2, except that Equation 1 was fit using either the full set of monomers, or only single-domain proteins.

|  | **Full set** $(A_{monomer} = 3.88 M^{0.785})$ | **Single-domain proteins** $(A_{monomer} = 5.82 M^{0.741})$ |
|---|---|---|
| **GNM** | 0.76 | 0.79 |
| **RMSF** | 0.81 | 0.84 |
| **Lindemann** | 0.78 | 0.73 |
| **RCI** | 0.73 | 0.67 |

**Table S3.** Pearson correlation coefficients between $A_{rel}$ values and different measures of intrinsic flexibility, as in Figure 2, for single-domain and two-domain proteins.

|  | **Single-domain proteins** | **Two-domain proteins** |
|---|---|---|
| **GNM** | 0.74 | 0.76 |
| **RMSF** | 0.82 | 0.81 |
| **Lindemann** | 0.76 | 0.81 |
| **RCI** | 0.60 | 0.86 |

**Table S4.** Pearson correlation coefficients between $A_{rel}$ values and intrinsic disorder predictions for the full set of monomeric crystal structures.

| Disorder predictor | r | P-value |
|---|---|---|
| ESpritz[2] | 0.043 | 0.0005 |
| FoldIndex[3] | 0.077 | 4 x 10$^{-10}$ |
| IUPred[4] | 0.006 | 0.6 |
| PreDisorder[5] | 0.13 | < 2.2 x 10$^{-16}$ |
| VSL2B[6,7] | 0.17 | < 2.2 x 10$^{-16}$ |

The Espritz predictor was used with all default parameters and the DisProt training set (the X-ray and NMR training sets gave even weaker correlations with $A_{rel}$). The global FoldIndex disorder prediction score was used, but the above was inverted, as a more negative FoldIndex score indicates a stronger disorder prediction. All other disorder predictors were used with default parameters, and the average disorder predictions were averaged over all residues. Only residues observed in the crystal structures were used in the disorder predictions.

**Table S5.** Pearson correlation coefficients between $A_{rel}$ values and different measures of intrinsic flexibility, as in Figure 2, for different SCOP structural classes.

|  | α | α+β | α/β | β |
|---|---|---|---|---|
| GNM | 0.76 | 0.76 | 0.77 | 0.75 |
| RMSF | 0.85 | 0.83 | 0.89 | 0.88 |
| Lindemann | 0.87 | 0.71 | 0.66 | 0.59 |
| RCI | 0.46 | 0.58 | 0.66 | 0.52 |

**Table S6.** Mean Pearson correlation coefficients between different measures of local flexibility.

|  | GNM | RMSF | RCI |
|---|---|---|---|
| $A_{rel}$ | 0.74 | 0.70 | 0.61 |
| GNM | - | 0.74 | 0.57 |
| RMSF | - | - | 0.61 |

# Supplementary References